\documentclass[cite1,clbiba,12pt,psfig]{article}
\usepackage{epsfig}
\usepackage{float}
\usepackage{graphicx}
\usepackage{overcite}
\usepackage{chapterbib}

\usepackage{fancyhdr}
\usepackage{hyperref}
\usepackage{authblk}

\title{Dielectric Properties of Water: A Molecular Dynamics Study on the Effects of Molecule Count and Cutoff Radius.  }
\date{}
\author[1]{\small Ra\'ul Fuentes-Azcatl \\
  \href{mailto:rfuentes@ifuap.buap.mx}{rfuentes@ifuap.buap.mx}}

\affil[1]
{\footnotesize Instituto de F\'isica ''Luis Rivera Terrazas'', Benem\'erita Universidad Aut\'onoma de Puebla, Apdo. Postal J-48, Puebla, 72570, M\'exico\\}

\newpage 

\begin{document}

\newpage

\maketitle
\begin{abstract}

This study investigates the dielectric properties of H$_2$O in relation to both the number of molecules and the cutoff radius within the simulation cell. We employ the flexible FBA/$\epsilon$ water model, known for its improved accuracy in reproducing water's properties across various thermodynamic states. We range from 60 to 1372 molecules and use cutoff radius from 6 nm to 1.2 nm, encompassing a wide range of calculations to assess their impact on the simulation results. These evaluations extend to critical properties like the dielectric constant, density, and phase change enthalpy.

\end{abstract}


\section{Introduction}


The utilization of interaction cutoffs between atoms is a common practice in molecular simulations. While it can introduce undesired effects, it becomes particularly valuable when dealing with a limited number of molecules. Through their interactions and the principles of statistical mechanics, these cutoffs enable the calculation of macroscopic properties, such as density and dielectric constants.

In the context of Lennard-Jones interactions, there is a minor energy impact due to their short-range nature. However, they can significantly affect pressure. Fortunately, for systems comprised of elements with shared characteristics, these effects can be analytically corrected.\cite{Allen1987,tuckerman2010statistical}

Long-range interactions can create important constraints on system size in theoretical-computational simulations of condensed matter under periodic boundary conditions.\cite{tuckerman2010statistical, Berendsen1981} The effects of electrostatic interactions described by the Coulomb potential have generally been established correctly \cite{tuckerman2010statistical}. Hydrodynamics can also be described through long-range interactions, as demonstrated by the description of the Oseen tensor.\cite{Jorgensen} Such hydrodynamic interactions can in fact cause significant finite-size effects, as demonstrated Dünweg and Kremer \cite{soper1997site}.

In various studies involving the confinement of water or other molecules \cite{Dillenburg2023,Mend,KOHLER201954,co2eGraf,VALENCIAORTEGA2019243}, a limited number of  molecules are often employed. However, when the research aims to analyze specific areas containing even fewer molecules, it raises reasonable doubts. Therefore, this study holds significance in establishing criteria for determining the minimum number of molecules required to conduct molecular dynamics investigations under conditions where a scarcity of molecules in certain regions is of interest.

Electrostatic interactions are typically calculated with a cutoff radius of at least 1 nm, a choice supported by experimental radial distribution functions \cite{soper1997site,tip4pe,spce,fbae,tip4pef} that reveal no significant features beyond 0.9 nm, as presented in this study. However, it's important to note that the absence of distinctive features doesn't imply the absence of liquid structure beyond this distance.

Theoretical computer simulations have demonstrated substantial ordering in water molecules up to a molecular separation of approximately 1.4 nm \cite{10.1063/1.468222}. Examining the time evolution of a finite dependent system is crucial at the cutoff point in order to recognize that this effect is not adequately addressed in the calculation of the dielectric properties of simulated water.
In this study, we conduct 100.0 ns simulations of water systems comprising 60 to 1372 molecules. We employ the FBA/$\epsilon$ \cite{fbae}water model, known for its accurate representation of various properties of water across different thermodynamic states. These simulations utilize cutoff radius ranging from 0.6 nm to 1.2 nm, allowing us to investigate the impact of system size.

Our primary goal is to quantify the influence of system size and cutoff radius on the density and dielectric properties under these conditions.

The remaining of the  paper goes as follows. In section II  
the models are introduced. Section III
shows the simulation details in Section IV 
 the results are analysed in Section V and finally  Conclusions are presented.

\subsection{The water Models}

The force field,FF, used in this work for water molecule is the recent FAB/$\epsilon$ \cite{fbae}, this FF is of three sites, on each atom there is a puntual charge and a Lennard-Jones site on Oxygen, the  model improves the reproduction of various thermodynamic properties at different pressure and temperature conditions of all three-site models, the model in addition has  a harmonic potential at the O-H$_{bond}$ and another at the angle formed by the three atoms forming the molecule. The table \ref{table2} shows the force field values.

\begin{table}
\caption{Parameters of the three-site water models considered in this work.
}
\label{table2}
\begin{tabular}{|ccccccccc|}
\hline\hline
model	&	$k_{b}$	&	$r _{OH}$ 	&	$k_{a}$	&	$\Theta$ 	&	$\varepsilon_{OO}$	&	$\sigma_{OO}$	&	$q_{O}$	&	$q_{H}$	\\

	&	kJ/ $mol$ {\AA}$^{2}$ 	&	{\AA}	&	kJ/ $mol$ rad$^{2}$	&	deg	&	kJ/mol	
	&	{\AA}	&	e	&	e	\\
	
FBA/$\epsilon $	\cite{fbae}&	3000	&	1.027	&	383	&	114.70	&	0.792324	&	3.1776	&	-0.8450	&	0.4225	\\

\hline
\end{tabular}
\end{table}

\newpage

 For the intermolecular potential between two molecules the LJ
 and Coulomb interactions are used for non-polarizable models,

\begin{equation}
\label{ff}
u(r) = 4\epsilon_{\alpha \beta} 
\left[\left(\frac {\sigma_{\alpha \beta}}{r}\right)^{12}-
  \left (\frac{\sigma_{\alpha \beta}}{r}\right)^6\right] +
\frac{1}{4\pi\epsilon_0}\frac{q_{\alpha} q_{\beta}}{r}
\end{equation}

\noindent where $r$ is the distance between sites $\alpha$ and $\beta$,
$q_\alpha$ is the electric charge of site $\alpha$, $\epsilon_0$ is the
permitivity of vacuum,  $\epsilon_{\alpha \beta}$ is the LJ energy scale
and  $\sigma_{\alpha \beta}$ the repulsive diameter for an $\alpha-\beta$ pair.
The cross interactions between unlike atoms are obtained using
the Lorentz-Berthelot mixing rules,

\begin{equation}
\label{lb}
\sigma_{\alpha\beta} = \left(\frac{\sigma_{\alpha\alpha} +
  \sigma_{\beta\beta} }{2}\right);\hspace{1.0cm} \epsilon_{\alpha\beta} =
\left(\epsilon_{\alpha\alpha} \epsilon_{\beta\beta}\right)^{1/2}
\end{equation}

In the case of the FAB/$\epsilon$ have the addition of armonical potential in O-H$_{bond}$ and in the angle H-O-H, as is indicated in the equation \ref{k}  and \ref{theta}.
 \begin{equation}
\label{k}
U_k(r)=\frac{k_r}{2}(r-r_0 )^2 
\end{equation}
 \begin{equation}
\label{theta}
U_{\theta}(\theta)=\frac{k_{\theta}}{2}(\theta-\theta_0)^2 ,
\end{equation}
\noindent where $r$ is the bond distance and $\theta$ is the bond 
angle. The subscript $0$ denotes their equilibrium
 values, $k_r$ and $k_{\theta}$ are the corresponding spring constants. 

\section {Simulation details}

Theoretical computational calculations were conducted by solving the equations of motion through molecular dynamics (MD) simulations in the liquid phase. These simulations were carried out using the isothermal-isobaric (NPT) ensemble, as described in the Tuckerman work \cite{tuckerman2010statistical}, to determine key properties such as density, the dielectric constant ($\epsilon$), and the heat of vaporization ($\Delta H$) of the system.

For these simulations, we utilized Gromacs package version 2022 \cite{Van_Der_Spoel2005-oo}. In our approach, molecules were initially placed randomly within the simulation cell, which was divided into slices of similar proportions along each axis. This division created planes within each axis, and at the intersection of these planes, molecules were positioned based on their center of mass. This process was implemented using custom-made programs.

We solved the leapfrog equations of motion \cite{tuckerman2010statistical} using a time step of 1 fs, because the model is flexible. Periodic boundary conditions were applied in all directions. It's important to note that the cutoff distance, which is a variable in this study, is reported in the results section. This cutoff distance remained consistent for both the real part of the Coulomb potential and the Lennard-Jones (LJ) interactions. Additionally, we applied analytical long-range corrections to the LJ interactions.

For the evaluation of reciprocal contributions, we employed the Particle Mesh Ewald (PME) method \cite{1995JChPh.103.8577E} with a grid spacing of 0.34 nm for the reciprocal vectors, using a spline of order 4. To control the temperature and pressure, we implemented the Nos\'e–Hoover thermostat \cite{10.1063/1.1378321} and Parrinello–Rahman barostat \cite{tuckerman2010statistical} with coupling times of 0.6 and 1.0 ps, respectively.

The final results were obtained after production runs lasting 100 ns, following an initial equilibration period of 1 ns during which the system consisted of randomly arranged molecules in the liquid phase.

\section {Results}
Systematically, the study began with 60 molecules and gradually increased to 1372 molecules, using the FBA/$\epsilon$ water model. Table \ref{table3} provides information on the number of molecules, initial cell size, and the cutoff radius employed. The criterion for selecting the initial cutoff radius was to ensure it was at least half the size of the simulation cell, in order to mitigate finite size issues.

It is essential to examine the behavior of the barostat since, in an unconstrained volume, the barostat redistributes energy to maintain system pressure. Therefore, initiating this analysis is crucial. The figure \ref{FigP} illustrates the pressure's behavior concerning both the cutoff radius and the number of molecules.

\begin{table}
\caption{Simulation data for different molecules count and cutoff radius}\

\label{table3}
\begin{tabular}{|cccccc|}
\hline\hline

H$_2$O	&	cell size	&	Volume	&	rcut	&	Pressure	&	Temperature	\\
molecules	&	nm	&	Nm$3$	&	nm	&	bar	&	K	\\
	&	x=y=z	&		&		&		&		\\

	\hline\hline
	
60	&	1.22076	&	1.82984	&	0.6	&	0.18	&	293.64	\\
108	&	1.48474	&	3.29777	&	0.6	&	0.44	&	295.31	\\
108	&	1.48426	&	3.25346	&	0.7	&	1.07	&	295.28	\\
256	&	1.98815	&	7.79609	&	0.6	&	0.94	&	296.81	\\
256	&	1.97279	&	7.71422	&	0.7	&	1.38	&	296.82	\\
256	&	1.96841	&	7.69434	&	0.9	&	1.46	&	296.85	\\
500	&	2.48624	&	15.0791	&	0.7	&	-0.02	&	297.35	\\
500	&	2.47351	&	15.0347	&	0.9	&	0.70	&	297.40	\\
500	&	2.48219	&	15.0324	&	1	&	0.52	&	297.41	\\
500	&	2.48028	&	15.0326	&	1.1	&	1.57	&	297.46	\\
500	&	2.46974	&	15.0315	&	1.2	&	0.32	&	297.46	\\
864	&	3.50546	&	26.0352	&	0.7	&	0.53	&	297.57	\\
864	&	3.50546	&	25.9835	&	0.9	&	0.84	&	297.65	\\
864	&	3.50546	&	25.9807	&	1	&	0.48	&	297.66	\\
864	&	2.95969	&	25.9796	&	1.1	&	0.52	&	297.66	\\
864	&	2.97066	&	25.9813	&	1.2	&	1.00	&	297.67	\\
1372	&	3.45928	&	41.3634	&	0.7	&	0.99	&	297.70	\\
1372	&	4.08971	&	41.2649	&	0.9	&	1.14	&	297.78	\\
1372	&	4.08971	&	41.2632	&	1	&	1.01	&	297.78	\\
1372	&	3.45739	&	41.2604	&	1.1	&	0.76	&	297.78	\\
1372	&	3.45217	&	41.2595	&	1.2	&	0.62	&	297.79	\\

\hline
\end{tabular}
\end{table}

 \subsection{Presure, P.} 

\begin{figure}[H]
	\centering
	{\includegraphics[clip,width=12cm,angle=-90]{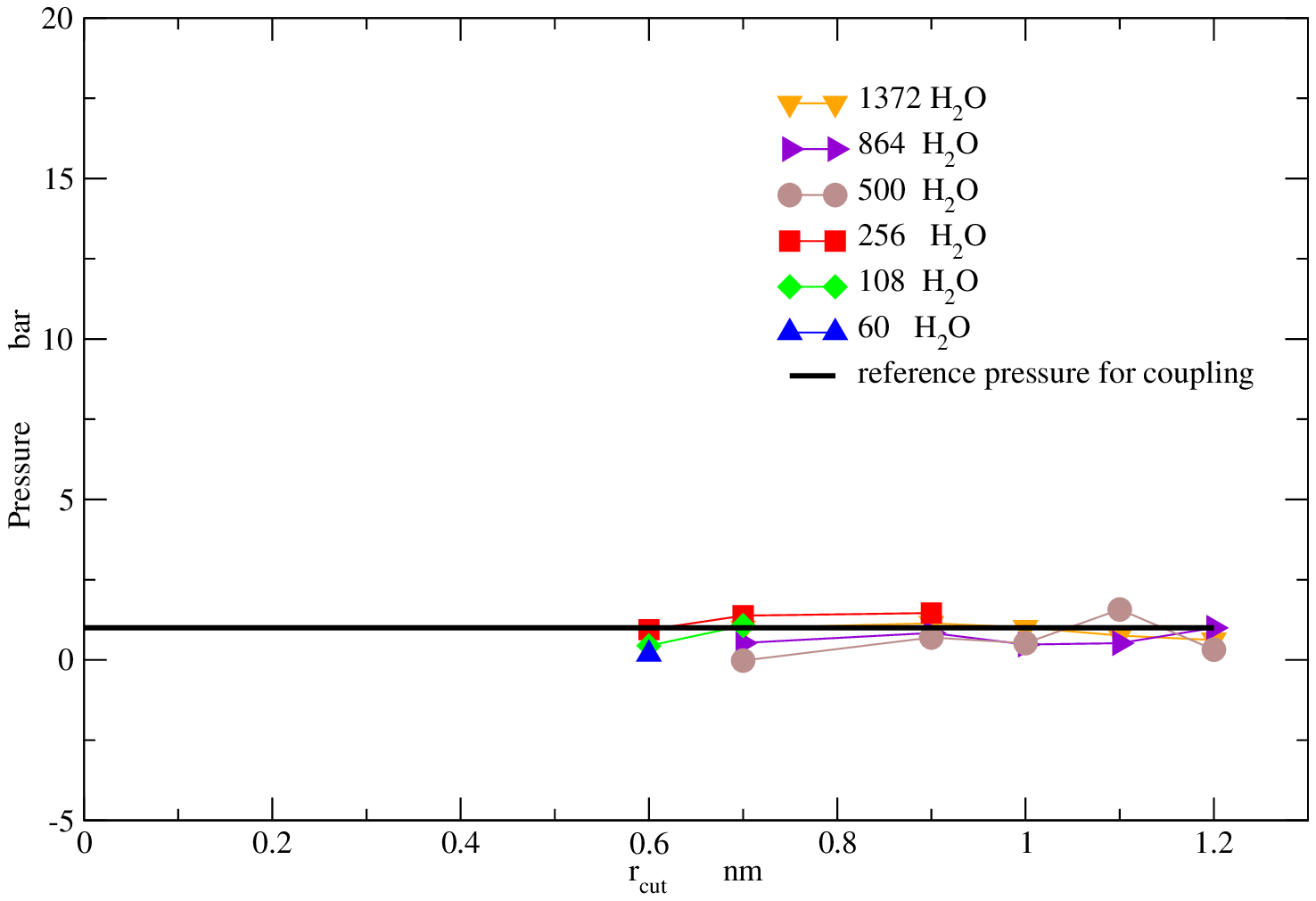}}
	
	\caption{Pressure as a function of r$_{cut}$ for different numbers of molecules. Reference pressure for coupling at 1 bar (black solid line), calculation with 60 molecules of water (blue triangle up), calculation with 108 molecules of water (green diamonds), calculation with 256 molecules of water (red squares), calculation with 500 molecules of water (brown circles), calculation with 864 molecules of water (violet triangle right), and calculation with 1372 molecules of water (orange triangle down).}
	\label{FigP}
\end{figure}

During the initial stages of a simulation, the system may have a different pressure than the desired one. The barostat applies a fictitious force to the particles in the simulation to gradually adjust the volume and, consequently, the pressure, to the desired value. In these calculations, pressure is controlled within the volume through the use of a barostat. Figure \ref{FigP} illustrates the variation of pressure as a function of the cutoff radius (r$_{cut}$) for different sets of molecules. It is evident that the pressure remains close to the average, and as the number of molecules increases, the pressure adjustment improves.

\subsection{Density, $\rho$.}

\begin{figure}[H]
	\centering
	{\includegraphics[clip,width=12cm,angle=-90]{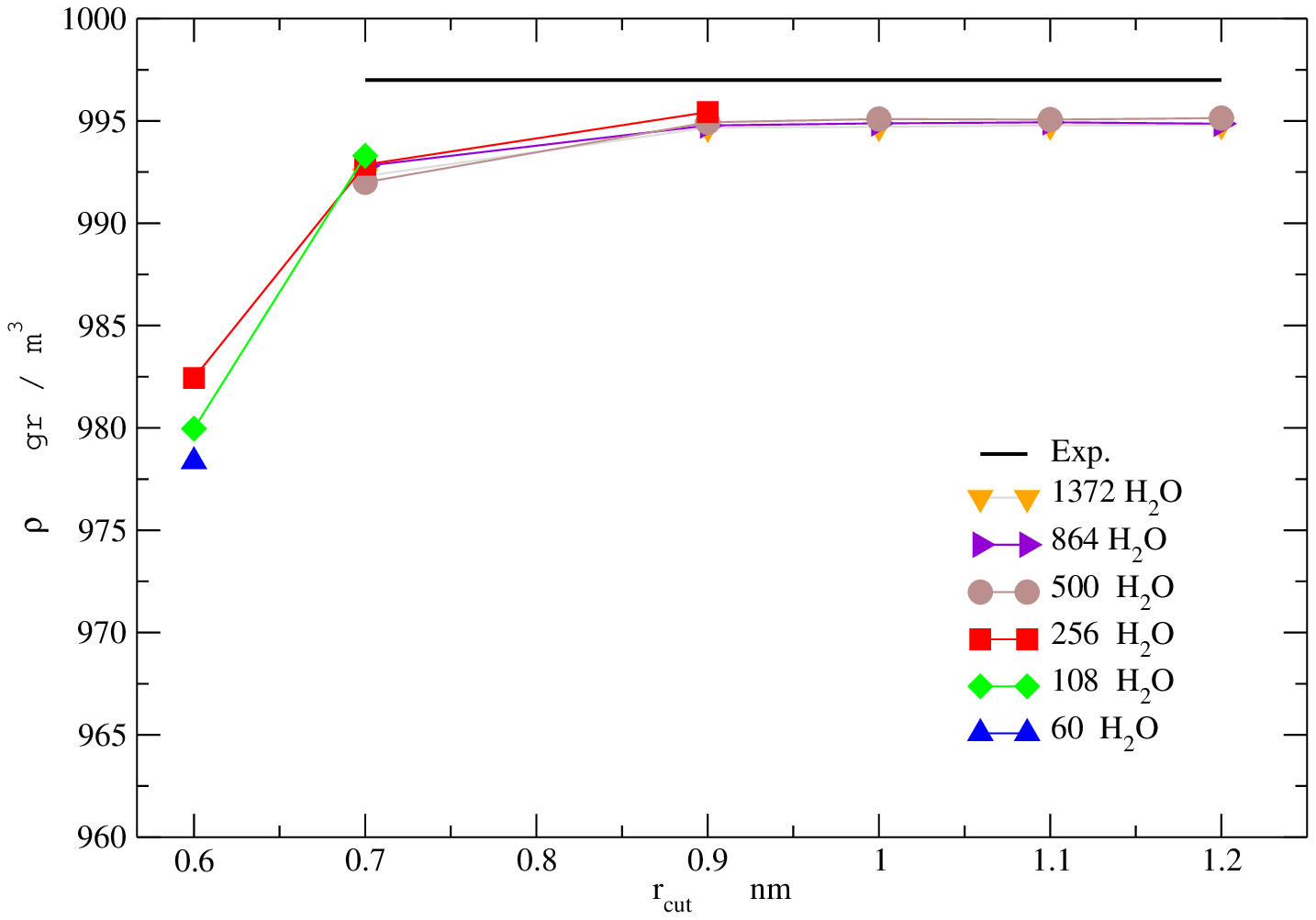}}
	
	\caption{Density as a function of r$_{cut}$ for different numbers of molecules at 1 bar and 298 K of pressure and temperature, respectively. Experimental data (black line)\cite{je60064a005}, calculation with 60 molecules of water (blue triangle up),  calculation with 108 molecules of water (green diamonds), calculation with 256 molecules of water (red squares), calculation with 500 molecules of water (brown circles), calculation with 864 molecules of water (violet triangle right), and calculation with 1372 molecules of water (orange triangle down).}
	\label{FigDens}
\end{figure}

The value reported by the FBA/$\epsilon$ model is 994 g/m$^3$, which exhibits a 0.3\% error compared to the experimental value at 1 bar of pressure and 298 K temperature. In Figure \ref{FigDens}, it is evident that this quantity can be reproduced with the same percentage error when starting with a cutoff radius of 0.9 nm and employing 256 molecules. The calculations remain consistent from 256 molecules up to 1372 molecules.

\subsection{Volume, V.} 

\begin{figure}[H]
	\centering
	{\includegraphics[clip,width=12cm,angle=-90]{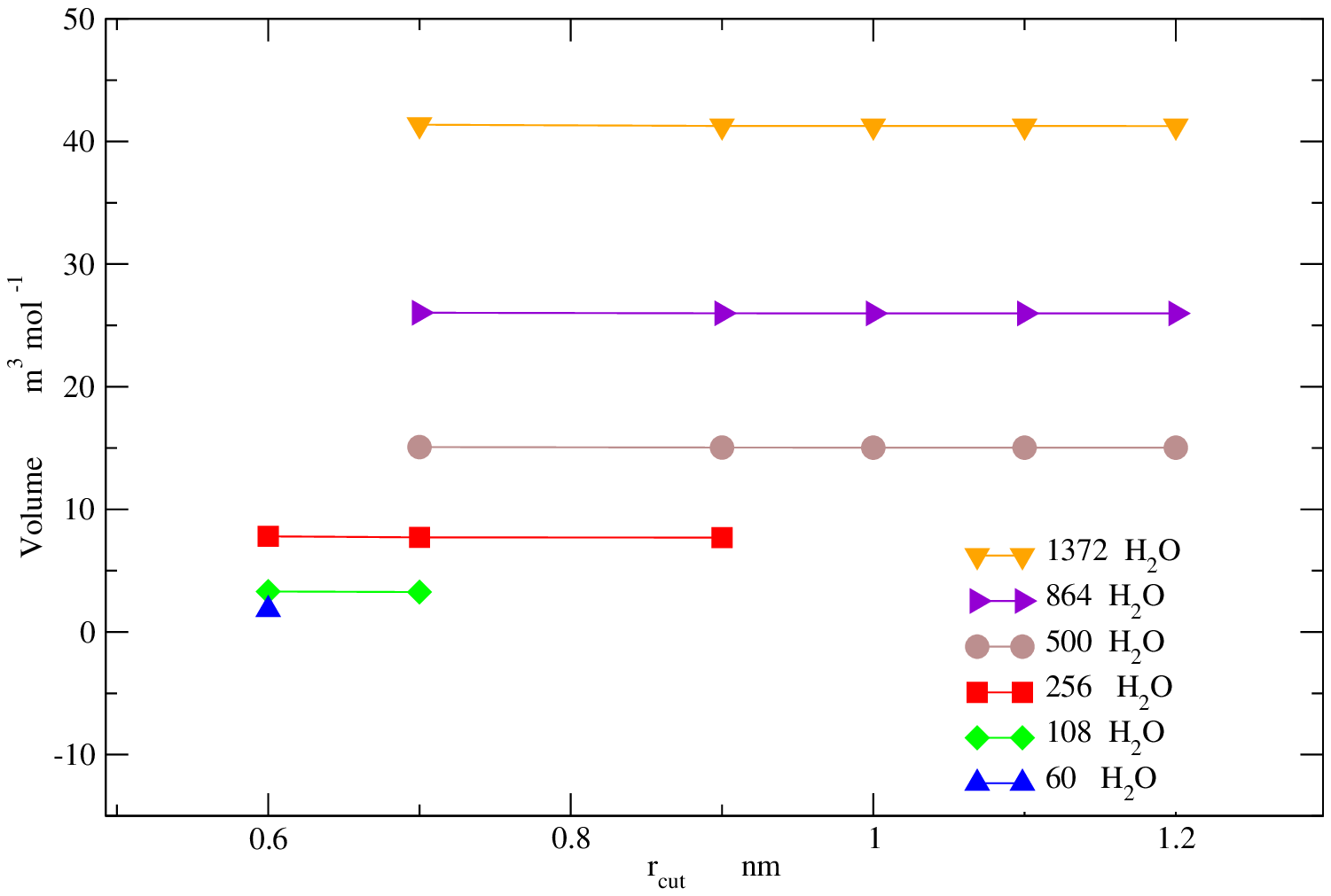}}
	
	\caption{The volume is plotted as a function of the cutoff radius (r$_{cut}$) for varying numbers of water molecules at a constant pressure of 1 bar and temperature of 298 K. The simulations encompass different scenarios: one involving 60 water molecules (indicated by blue triangle up), another with 108 water molecules (depicted by green diamonds), a third with 256 water molecules (represented by red squares), followed by one with 500 water molecules (denoted by brown circles), another with 864 water molecules (indicated by violet triangle right), and finally, a simulation with 1372 water molecules (highlighted by orange triangle down).}
	\label{FigV}
\end{figure}

In Figure \ref{FigV}, the behavior of the simulation cell volume is depicted, ranging from 60 molecules to 1372 molecules, with various cutoff radius. The volume increases as a result of the number of molecules used in the simulation cell and remains consistent for each set of molecules. Thus, for each molecule count, the volume remains constant, thanks to the barostat, which adjusts the energy to maintain the pressure within the reference value.

\subsection{The Heat of Vaporization, $\Delta H_{vap}$.}

\begin{figure}[H]
	\centering
	{\includegraphics[clip,width=12cm,angle=-90]{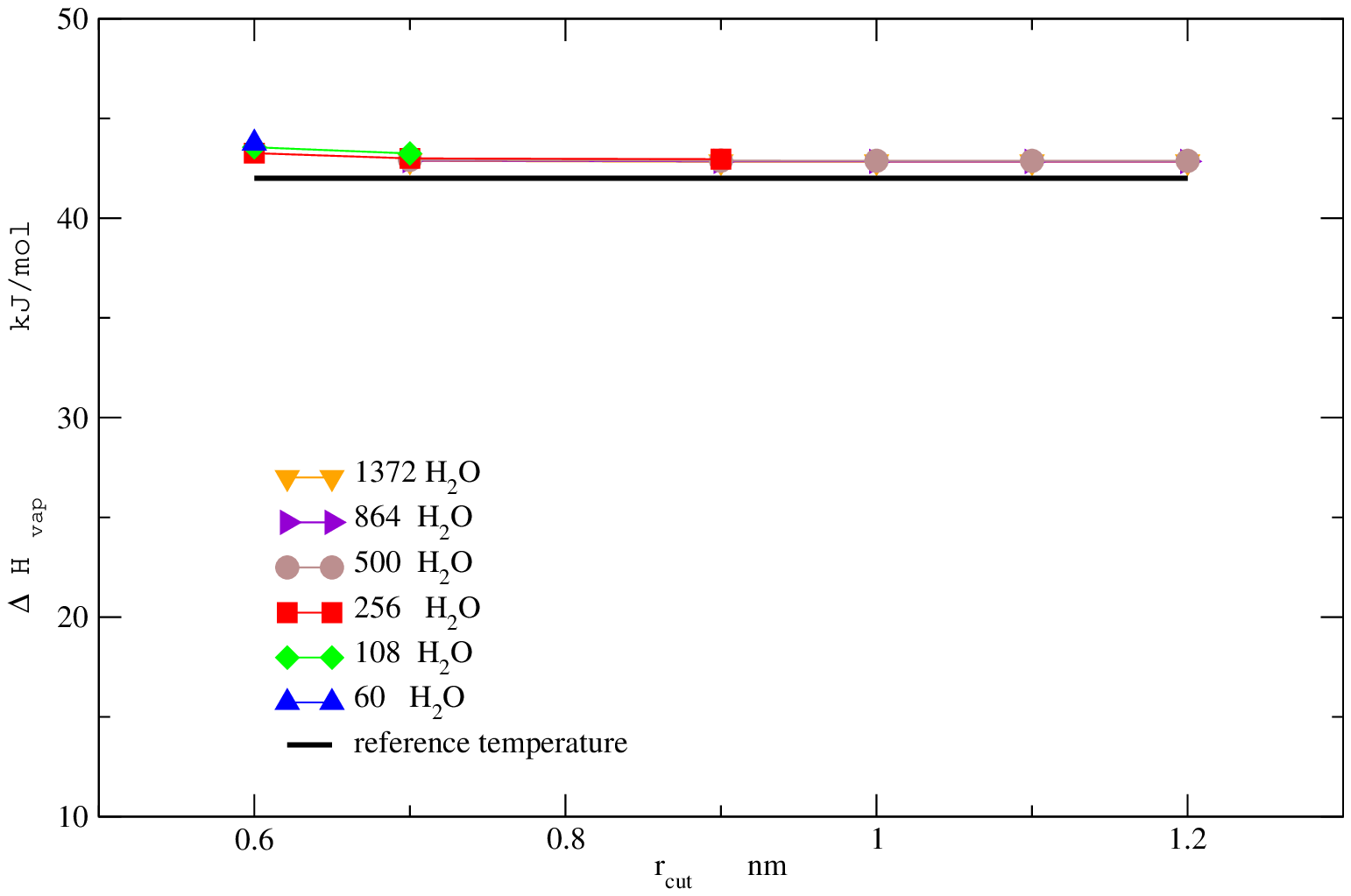}}
	
	\caption{Heat of vaporization $\Delta H_{vap}$ versus r$_{cut}$ for different number of molecules at 1bar and 298 K of pressure and temperature respectively. Experimental data (black line)\cite{10.1063/1.1461829}, calculation with 60 molecules of water (blue triangle up), calculation with 108 molecules of water(green diamonds ), calculation with 256 molecules of water(red squares), calculation with 500 molecules of water(brown circles), calculation with 864 molecules of water(violet triangle right), calculation with 1372 molecules of water(orange triangle down).}
	\label{FigDH}
\end{figure}

The value provided by the FBA/$\epsilon$ model is 42.95 kJ/mol, displaying a 2\% deviation when compared to the experimental value at a pressure of 1 bar and temperature of 298 K. In the calculations performed within this study, involving different quantities of molecules and varying cutoff radius, the figures presented in Figure \ref{FigDH} unmistakably indicate that the reproduced values remain in close agreement with those reported by Fuentes et al\cite{fbae}. This consistency holds true throughout the entire range of molecules that were both calculated and analyzed in the course of this research.

\subsection{Temperature, T.}

\begin{figure}[H]
	\centering
	{\includegraphics[clip,width=12cm,angle=-90]{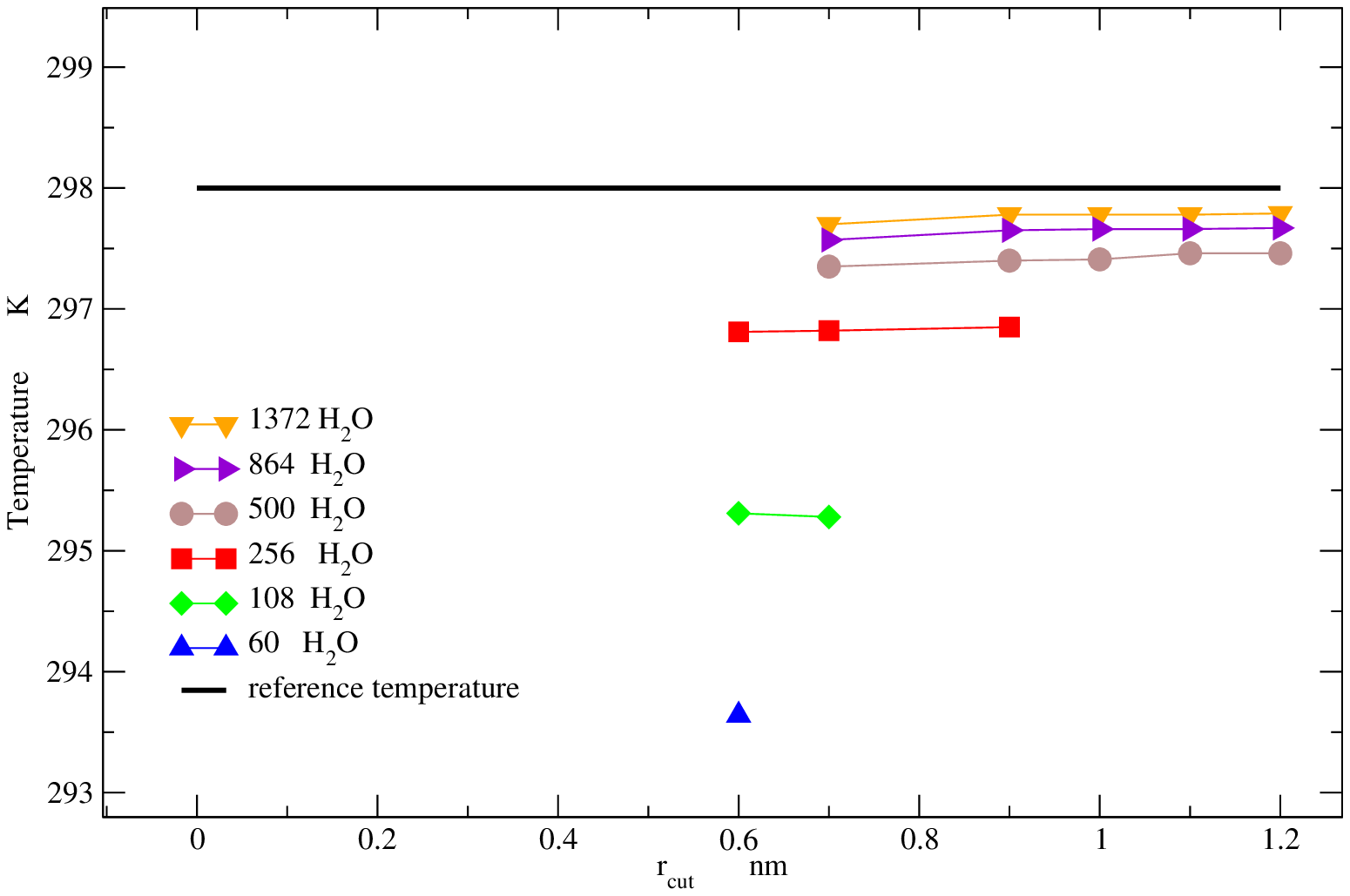}}
	
	\caption{Temperature versus r$_{cut}$ for different numbers of molecules at 1 bar of pressure. Calculations include 60 molecules of water (blue triangle up), 108 molecules of water (green diamonds), 256 molecules of water (red diamonds), 500 molecules of water (brown circles), 864 molecules of water (violet triangle right), and 1372 molecules of water (orange triangle down).}
	\label{FigTemp}
\end{figure}

As the barostat redistributes energy to maintain the reference pressure, the thermostat also redistributes energy to keep the temperature within the parameter set in the simulation. In Figure \ref{FigTemp}, the calculated temperature values are shown under different sets of molecules and for calculations performed with varying cutoff radius. It can be observed that the temperature remains constant for each molecule count, and as the system size increases, the thermostat adjusts the temperature more accurately, reaching an error of 0.1\% when using 1372 molecules. Additionally, we note that with 60 molecules, the error increases to 1\%, which is still a relatively low value.

\subsection{Radial distribution function, g(r).}

\begin{figure}[H]
	\centering
	{\includegraphics[clip,width=12cm,angle=-90]{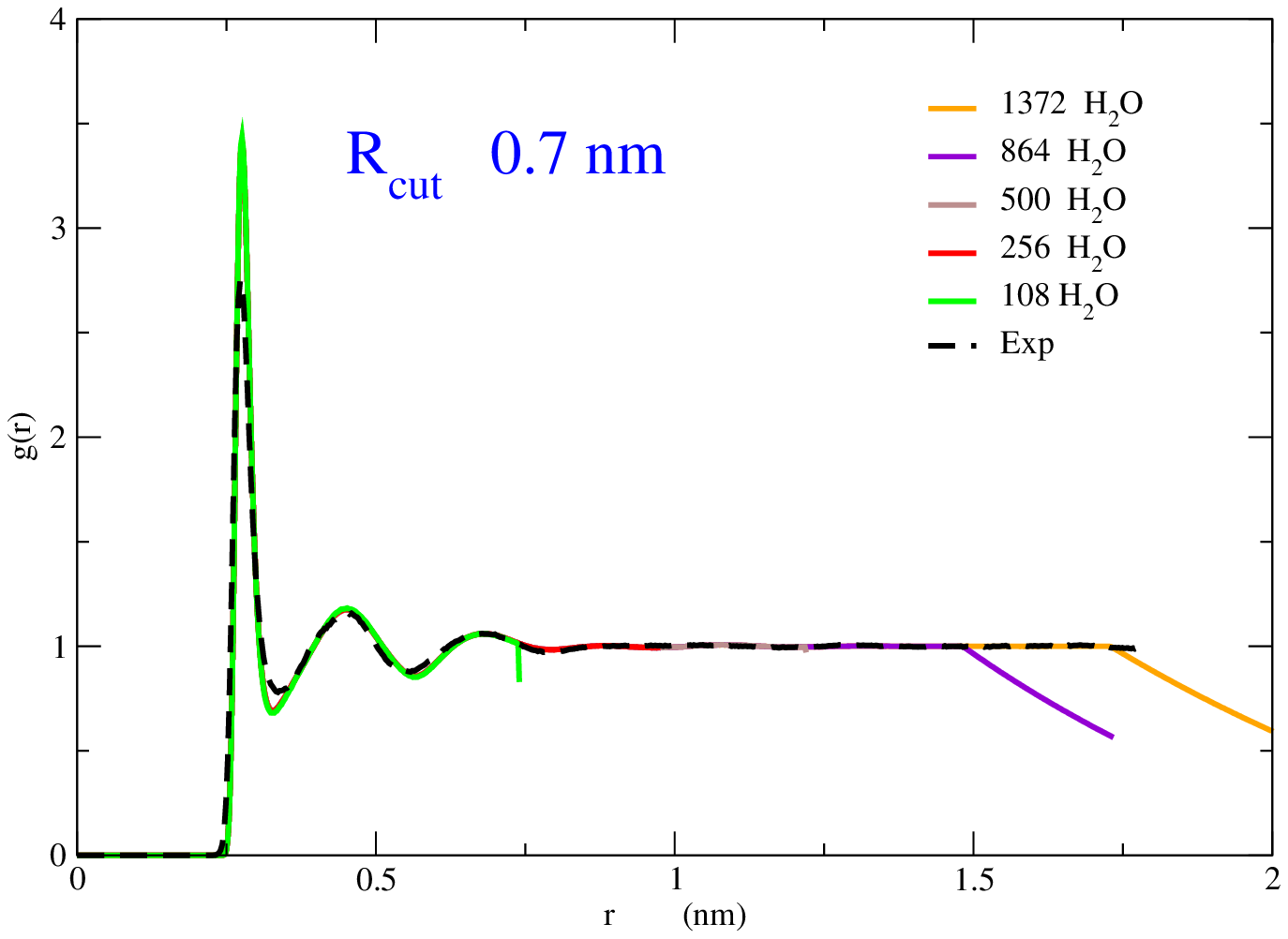}}
	
	\caption{Radial distribution, g(r), for different numbers of molecules at 1 bar and 298 K of pressure and temperature, respectively, using a cutoff radius (r$_{cut}$) of 0.7 nm. Experimental data (black line)\cite{2000CP}, calculation with 108 molecules of water (green line), calculation with 256 molecules of water (red line), calculation with 500 molecules of water (brown line), calculation with 864 molecules of water (violet line), and calculation with 1372 molecules of water (orange line).}
	\label{FigGr1}
\end{figure}

\begin{figure}[H]
	\centering
	{\includegraphics[clip,width=12cm,angle=-90]{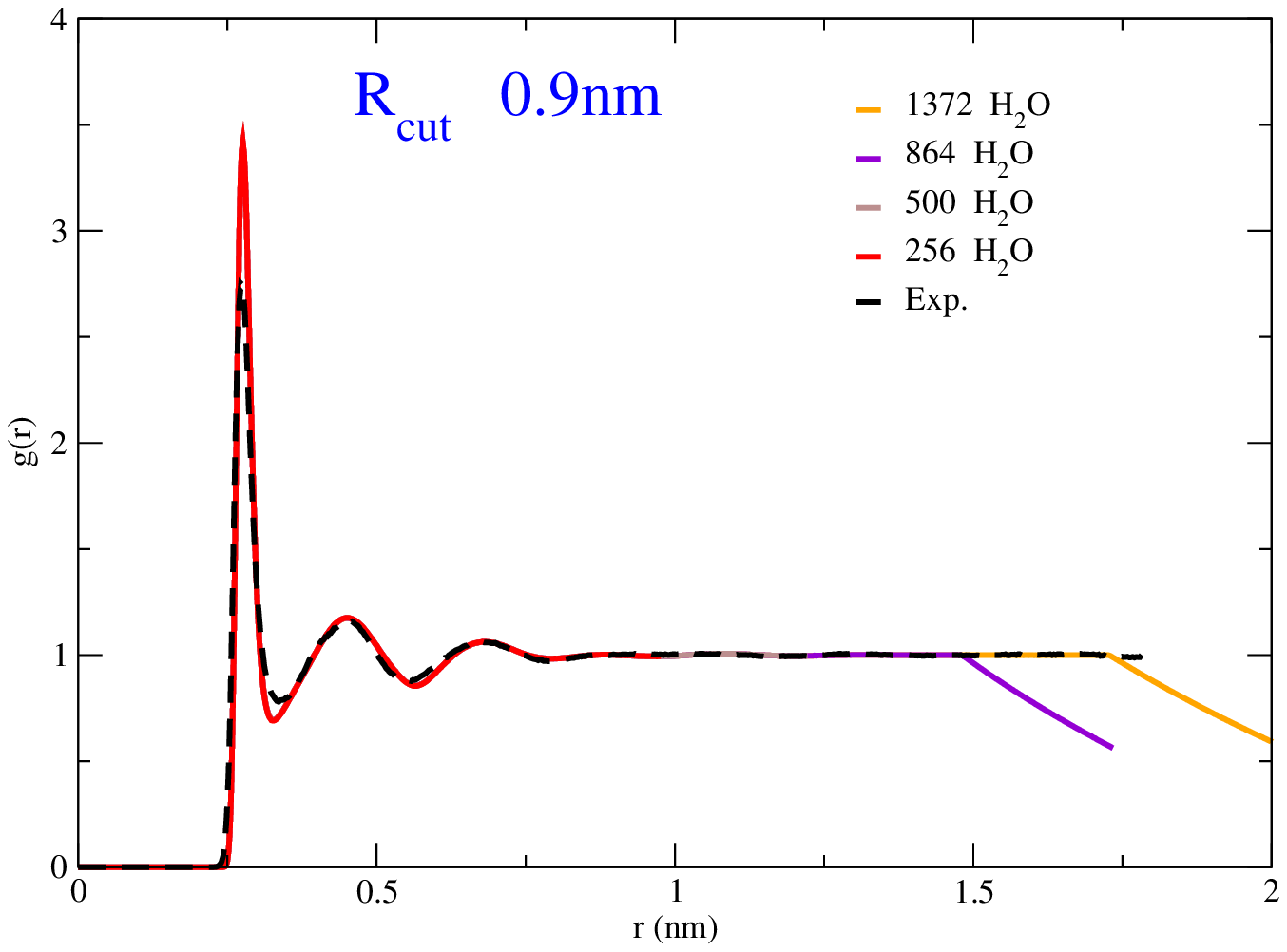}}
	
	\caption{Radial distribution, g(r), for different numbers of molecules at 1 bar and 298 K of pressure and temperature, respectively, using a cutoff radius (r$_{cut}$) of 0.7 nm. Experimental data (black line)\cite{2000CP}, calculation with 108 molecules of water (green line), calculation with 256 molecules of water (red line), calculation with 500 molecules of water (brown line), calculation with 864 molecules of water (violet line), and calculation with 1372 molecules of water (orange line).}
	\label{FigGr2}
\end{figure}

The structure of liquid water is analyzed using the radial distribution function, g(r). Figure \ref{FigGr1} and \ref{FigGr2} display the oxygen-oxygen pair distribution function obtained in NPT simulations at T = 298 K and P = 1 bar for liquid water. The results reveal a higher first peak compared to the experimental results\cite{} for the liquid. It is evident that with a cutoff radius of 0.7 nm, the function is truncated only when 108 molecules are used, while for a greater number of molecules, the function extends beyond 0.7 nm.

\subsection{Dielectric Properties}

\subsubsection{Dielectric constant $\epsilon$.}

\begin{figure}[H]
	\centering
	{\includegraphics[clip,width=12cm,angle=0]{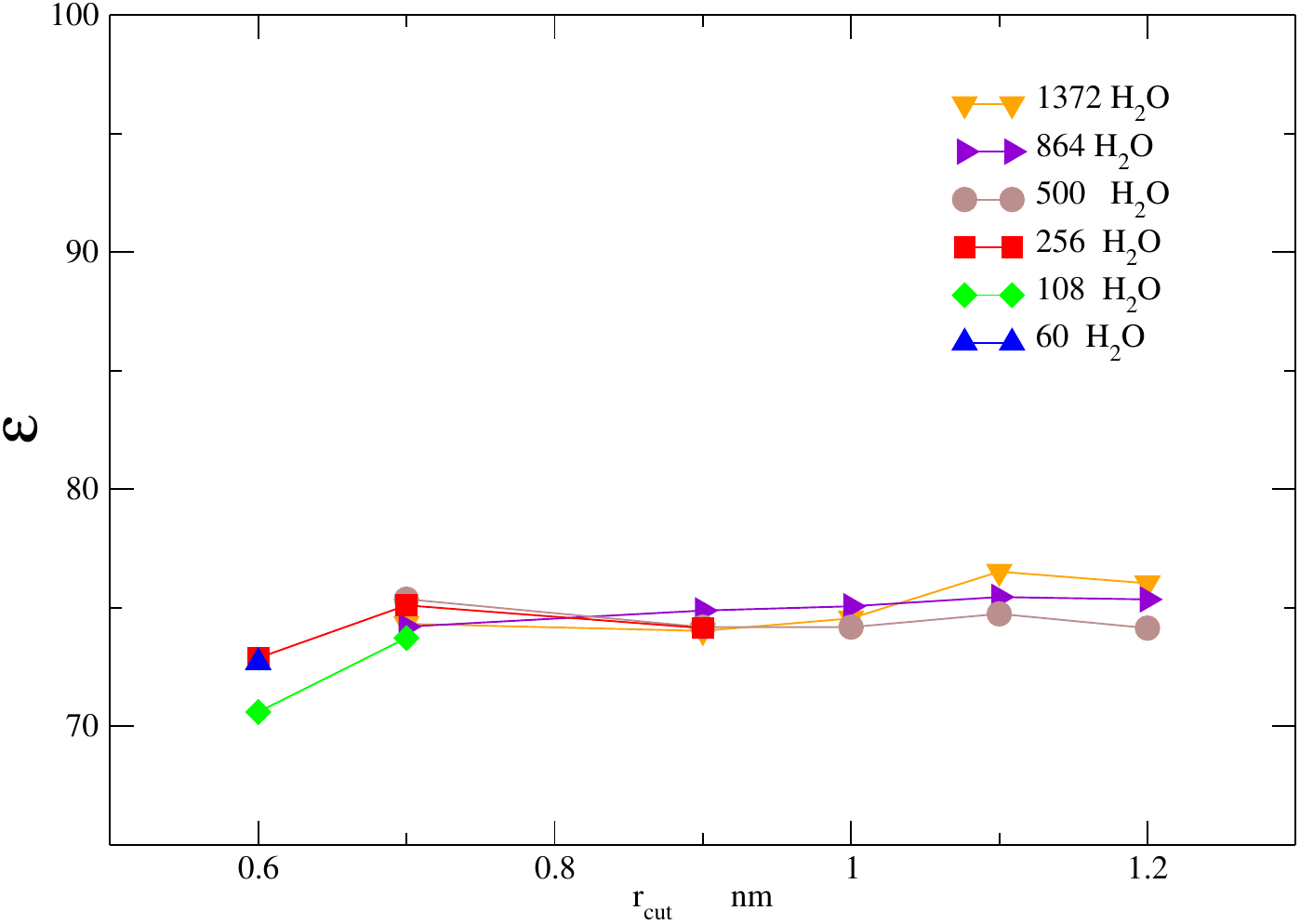}}
	
	\caption{Dielectric constant versus r$_{cut}$ for different number of molecules at 1bar and 298 K of pressure and temperature respectively, calculation with 60 molecules of water (blue triangle up), calculation with 108 molecules of water(green squares), calculation with 256 molecules of water(red diamonds), calculation with 500 molecules of water(brown circles), calculation with 864 molecules of water(violet triangle right), calculation with 1372 molecules of water(orange triangle down).}
	\label{FigCteDiel}
\end{figure}

The accurate determination of the dielectric constant requires extended simulations to ensure that the average dipole moment of the system hovers around zero\cite{tip4pe}. The dielectric constant results are presented in Figure \ref{FigCteDiel} for varying quantities of molecules and different cutoff radius at a temperature of 298 K and pressure of 1 bar.

The static dielectric constant of water, denoted as $\epsilon$, is a collective property of a set of water dipoles and can be computed from the fluctuations in the equilibrium total dipole moment, ($< M^2 >$ - $<M>^2$). The calculation of the dielectric constant\cite{Neumann} is derived using Equation \ref{diel} of the total dipole moment:

\begin{equation}
\label{diel}
\epsilon=1 + \frac{1}{3k_B TV \epsilon _0}(<M^2>-<M>^2),
\end{equation}

where $k_B$ is the Boltzmann constant, T is the absolute temperature, $\epsilon_0$ is the vacuum permittivity and V represents the volume.

The value reported by the FBA/$\epsilon$ model is $\epsilon$=75.65, representing a 3\% deviation from the experimental value of 78.5. As shown in the data presented in Figure \ref{FigCteDiel}, the values range from 74.5 to 76.3, spanning simulations with 108 molecules and a cutoff radius of 0.7 nm up to simulations with 1372 molecules and larger cutoff radius. A discrepancy with the experimental value is observed, ranging from 5\% for a cutoff radius of 0.7 nm and 108 molecules to 2.2\% for a cutoff radius of 1.1 nm and 1372 molecules. This highlights the importance of considering the limitations when analyzing systems with a small number of molecules.

\begin{figure}[H]
	\centering
	{\includegraphics[clip,width=12cm,angle=0]{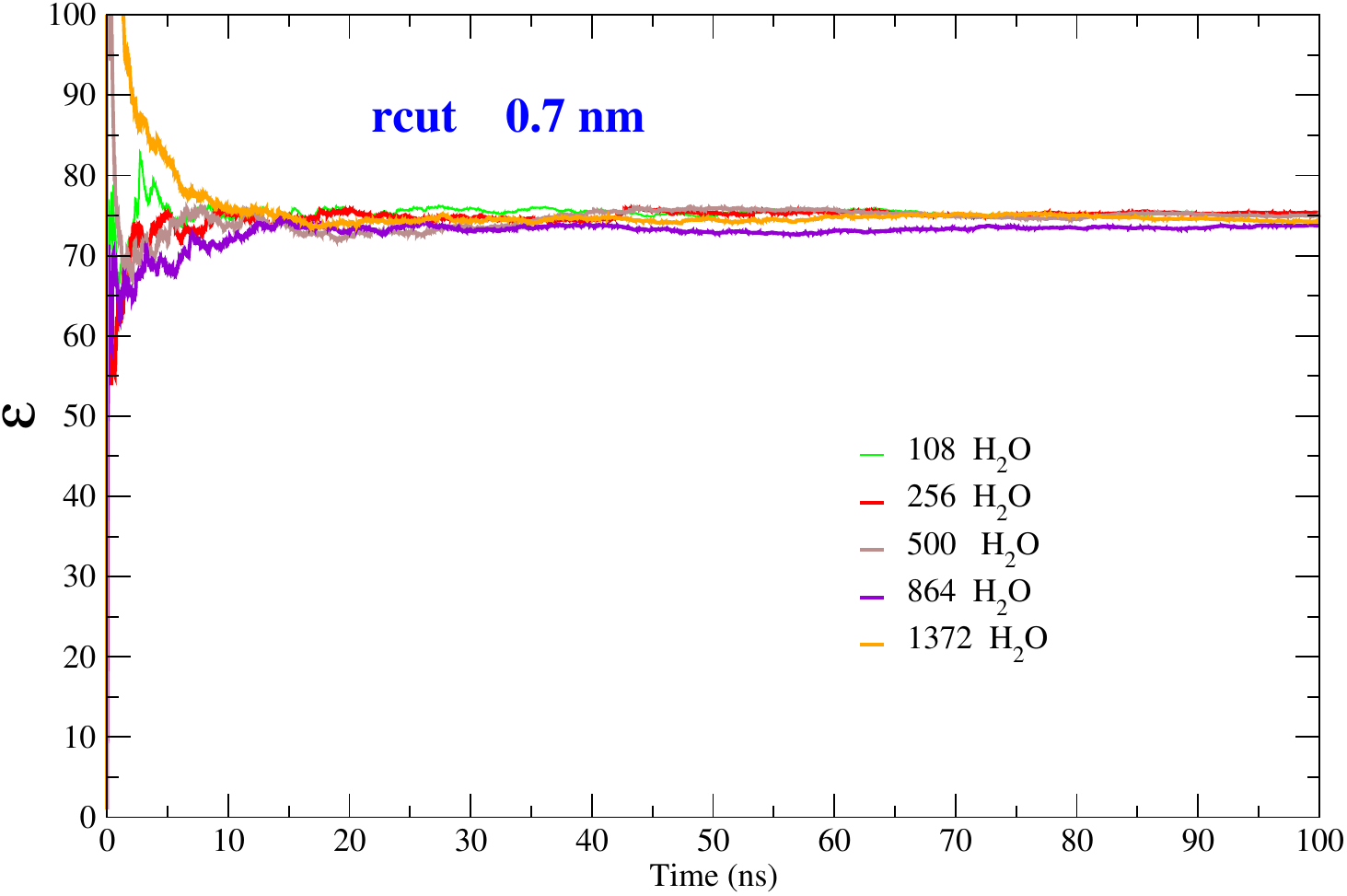}}
	
	\caption{Dielectric constant versus time with a r$_{cut}$=0.7 nm for different number of molecules at 1bar and 298 K of pressure and temperature respectively, calculation with 108 molecules of water(green line), calculation with 256 molecules of water(red line), calculation with 500 molecules of water(brown line), calculation with 864 molecules of water(violet line), calculation with 1372 molecules of water(orange line).}
	
	\label{Figr7Time}
\end{figure}

We can observe in Figure \ref{Figr7Time} the calculation of the dielectric constant over time with a cutoff radius of 0.7 nm. To obtain a reliable estimate of this property, simulations of at least 50 ns are required, as it stabilizes and converges to an average value after this time.

\begin{figure}[H]
	\centering
	{\includegraphics[clip,width=12cm,angle=-90]{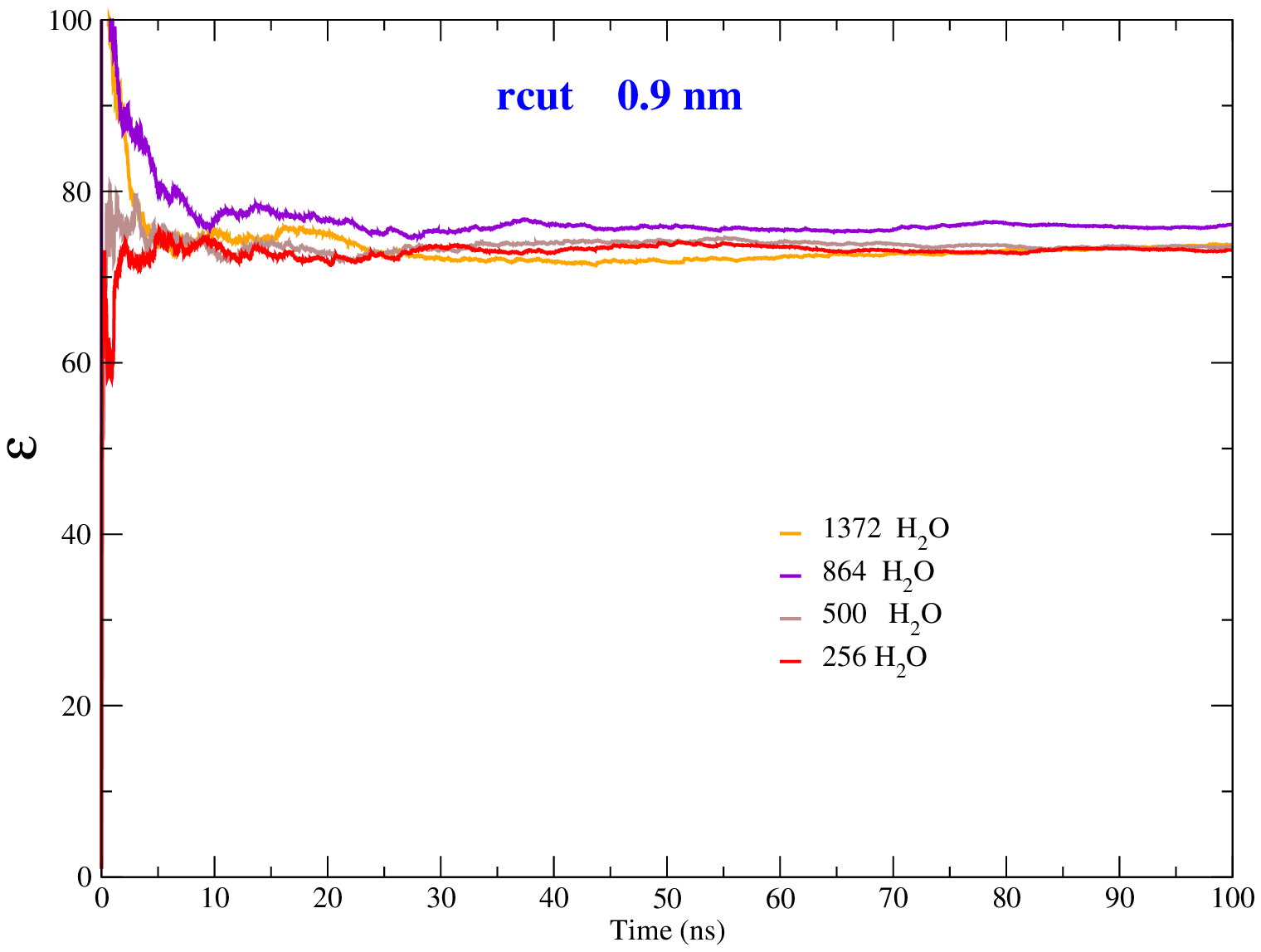}}
	
	\caption{Dielectric constant versus time with a r$_{cut}$=0.9 nm  for different number of molecules at 1bar and 298 K of pressure and temperature respectively, calculation with 108 molecules of water(green line), calculation with 256 molecules of water(red line), calculation with 500 molecules of water(brown line), calculation with 864 molecules of water(violet line), calculation with 1372 molecules of water(orange line).}

	\label{Figr9Time}
\end{figure}

In the case of using a cutoff radius of 0.9 nm, as shown in Figure \ref{Figr9Time}, it is noticeable that stability takes longer to achieve, typically requiring at least 80 ns for stabilization. Additionally, systematically, with 864 molecules, the value is 1.5\% higher than with other molecule quantities.

\subsubsection{Dipole moment, $\mu$.}

The distribution of the dipole moment is consistent with the size of the system, growing as the system expands. This observation indicates independence from the cutoff radius, as illustrated in the figure\ref{DM}.

	\begin{figure}[H]
	\centering
	{\includegraphics[clip,width=12cm,angle=-90]{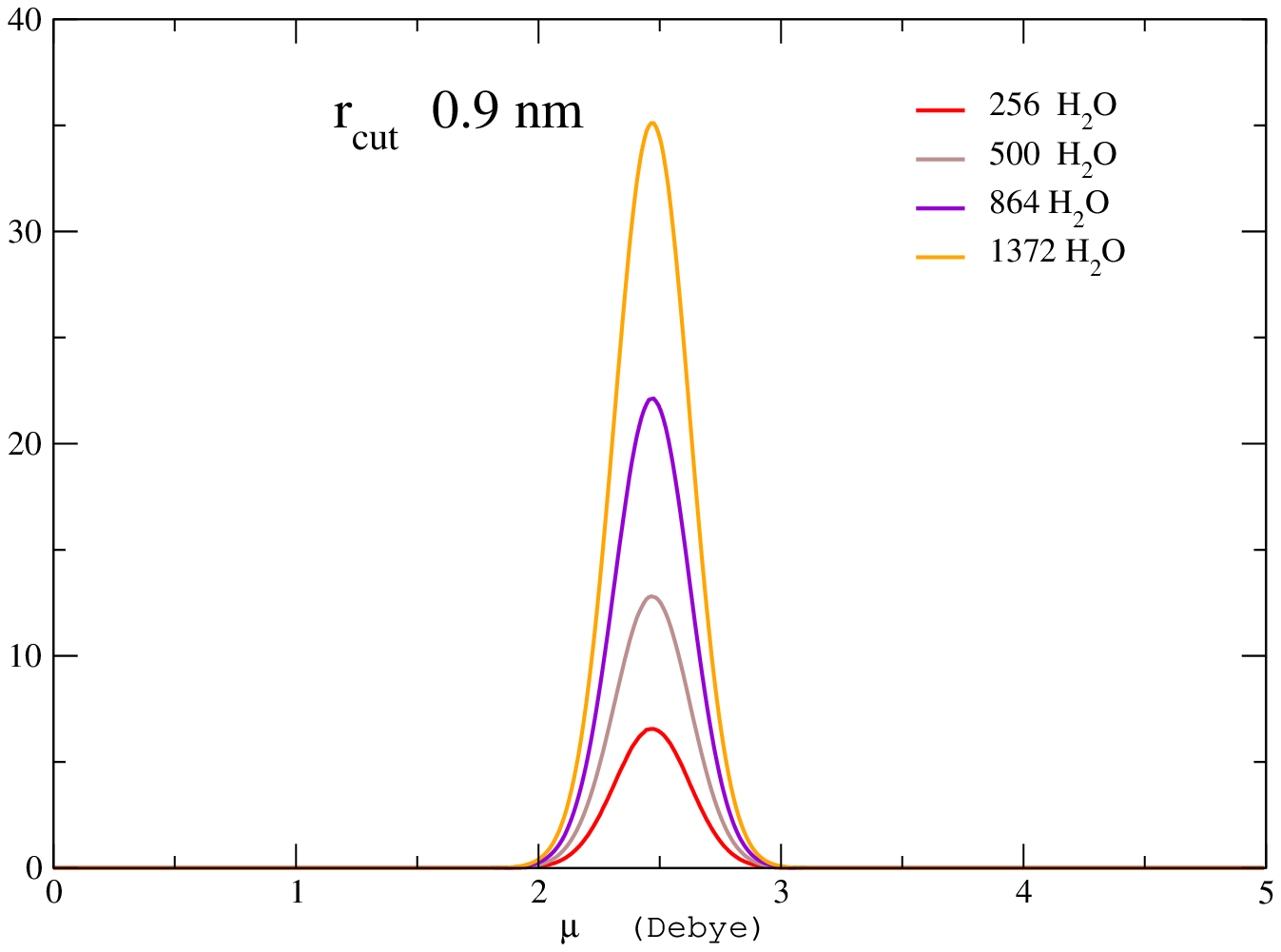}}
	
	\caption{Molecular dipole moment $\mu$ distributions for different numbers of molecules with a r$_{cut}$=0.9 nm, at 1 bar and 298 K of pressure and temperature, respectively. Calculations include 256 molecules of water (red line), 500 molecules of water (brown line), 864 molecules of water (violet line), and 1372 molecules of water (orange line).}
	\label{FigDM}
\end{figure}

\subsubsection{Finite system Kirkwood g-factor, G$_k$.}

The Kirkwood G-factor was assessed using the following approach: when the dipole $<M>$ is represented as a sum of N molecular moments $\mu$, it can be expressed as the equation \ref{gk1}, \cite{PhysRevLett.98.247401}
\begin{equation}
\label{gk1}
<M^2>=N \mu^2  (1+ \sum_{i}N_i <cos \theta_i>) = N \mu^2 G_K, 
\end{equation}
\\
the summation encompasses the coordination shells that surround a tagged molecule. Within this framework, N$_i$ signifies the number of molecules in the i-th coordination shell, and $<cos\theta_i>$ represents the average cosine of the angle formed between a dipole in the i-th coordination shell and the dipole of the central molecule. The ultimate expression defines the correlation factor G$_K$.

\begin{equation}
\label{gk2}
G_K= <{\bf M}^2> / N \mu^2  
\end{equation}
\\
N represents the total number of molecules, while M is the summation of all molecular dipole moments, denoted as $ {\mu} _i$, within the system (including the dipole of the initial molecule). Local orientational correlations tend to average out due to thermal motion after a few coordination shells.
	\begin{figure}[H]
	\centering
	{\includegraphics[clip,width=12cm,angle=-90]{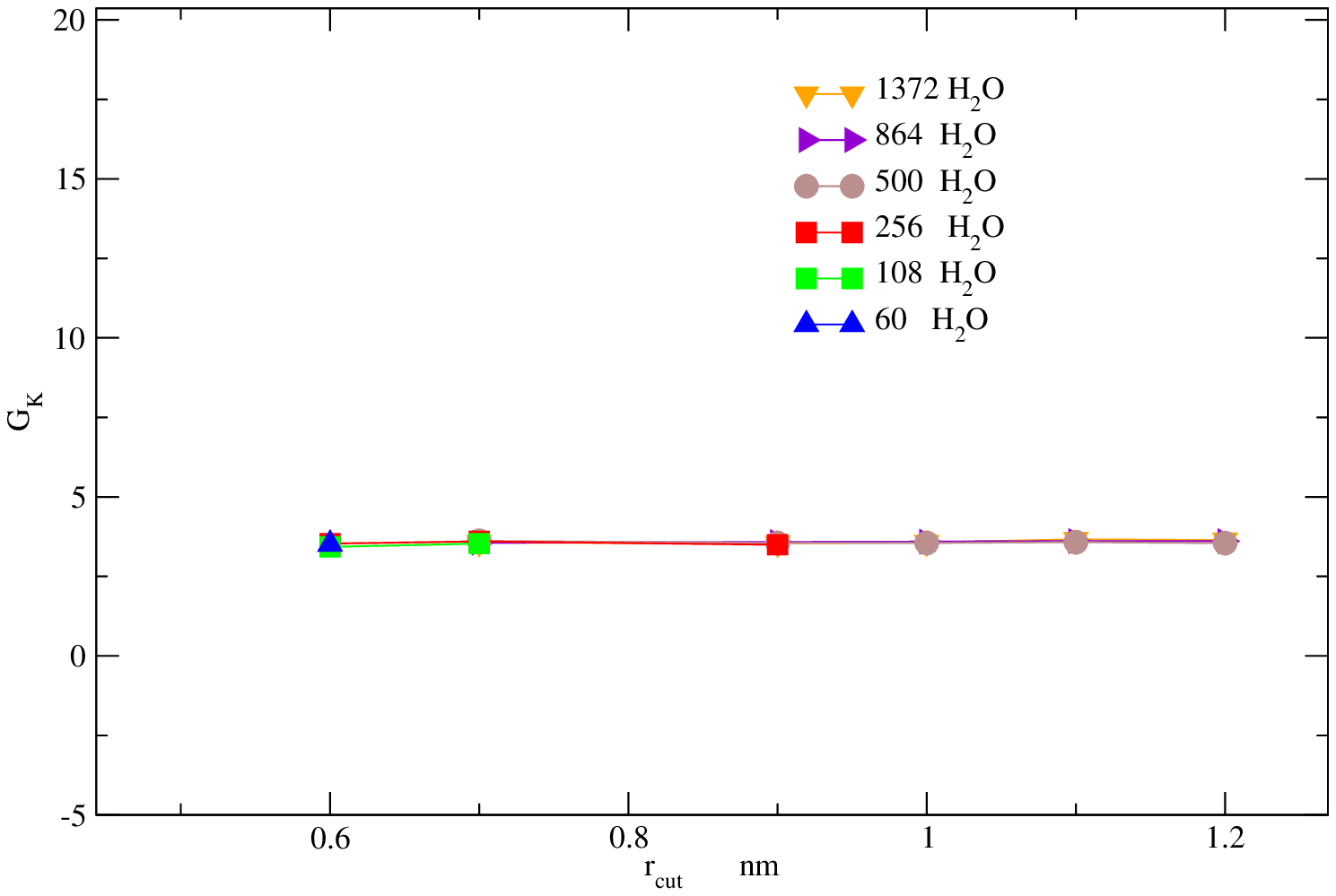}}
	
	\caption{ Finite system Kirkwood g-factor G$k$ versus r${cut}$ for different numbers of molecules at 1 bar and 298 K of pressure and temperature, respectively. Calculations include 60 molecules of water (blue triangle up), 108 molecules of water (green squares), 256 molecules of water (red diamonds), 500 molecules of water (brown circles), 864 molecules of water (violet triangle right), and 1372 molecules of water (orange triangle down).}
	\label{FigGk}
\end{figure}

\section {Conclusion}

This work represents a systematic study aimed at comprehending the influence of both molecule count and cutoff radius on density, pressure, and volume, and determining the extent of their impact. The dielectric properties consistently exhibit a stable trend starting from 256 molecules with a cutoff radius of 7 nm. This consistency is particularly significant in confinement studies where a small number of molecules are in proximity to layers. These results provide strong assurance that the dielectric values closely match experimental data with minimal error.

Furthermore, we note that the heat of vaporization, $\Delta$H$_{vap}$, remains constant regardless of the number of molecules and the cutoff radius. This observation suggests that the energy is accurately represented not only in smaller systems but also in those comprising a larger number of molecules.
Furthermore, we notice that both the polarization described by G$_k$ and the dipole moment distribution are also independent of the system size. Therefore, we can confidently conclude that conducting analyses with a reduced number of molecules, as is often the case in confined systems, is appropriate, given the values presented here.

\section {Acknowledgements}
 
The author extends their appreciation to CONAHCYT for this research and the SNI incentive.

\bibliography{achemso}

\providecommand{\latin}[1]{#1}
\makeatletter
\providecommand{\doi}
  {\begingroup\let\do\@makeother\dospecials
  \catcode`\{=1 \catcode`\}=2 \doi@aux}
\providecommand{\doi@aux}[1]{\endgroup\texttt{#1}}
\makeatother
\providecommand*\mcitethebibliography{\thebibliography}
\csname @ifundefined\endcsname{endmcitethebibliography}
  {\let\endmcitethebibliography\endthebibliography}{}
\begin{mcitethebibliography}{81}
\providecommand*\natexlab[1]{#1}
\providecommand*\mciteSetBstSublistMode[1]{}
\providecommand*\mciteSetBstMaxWidthForm[2]{}
\providecommand*\mciteBstWouldAddEndPuncttrue
  {\def\EndOfBibitem{\unskip.}}
\providecommand*\mciteBstWouldAddEndPunctfalse
  {\let\EndOfBibitem\relax}
\providecommand*\mciteSetBstMidEndSepPunct[3]{}
\providecommand*\mciteSetBstSublistLabelBeginEnd[3]{}
\providecommand*\EndOfBibitem{}
\mciteSetBstSublistMode{f}
\mciteSetBstMaxWidthForm{subitem}{(\alph{mcitesubitemcount})}
\mciteSetBstSublistLabelBeginEnd
  {\mcitemaxwidthsubitemform\space}
  {\relax}
  {\relax}


\bibitem{Allen1987}Allen, M. \& Tildesley, D. Computer Simulations of Liquids. {\em Oxford Science}. (1987)

\bibitem{tuckerman2010statistical}Tuckerman, M. Statistical mechanics: theory and molecular simulation. (Oxford university press,2010)
\mciteBstWouldAddEndPuncttrue
\mciteSetBstMidEndSepPunct{\mcitedefaultmidpunct}
{\mcitedefaultendpunct}{\mcitedefaultseppunct}\relax
\EndOfBibitem
\bibitem{Berendsen1981}Berendsen, H., Postma, J., Van Gunsteren, W. \& Hermans, J. Intermolecular Forces.  (1981)
\bibitem{Jorgensen}Jorgensen,W. L., ChandrasekharJ., Madura, J. D.  \& Klein, M. L. J. Chem. Phys.79, 926  (1983)

\bibitem{soper1997site}Soper, A., Bruni, F. \& Ricci, M. Site–site pair correlation functions of water from 25 to 400 C: Revised analysis of new and old diffraction data. {\em The Journal Of Chemical Physics}. \textbf{106}, 247-254 (1997)
\bibitem{Dillenburg2023}F. Dillenburg, R., Abal, J. \& Barbosa, M. Computational Investigation on Water and Ion Transport in MoS2 Nanoporous Membranes: Implications for Water Desalination. {\em ACS Applied Nano Materials}. \textbf{6}, 4465-4476 (2023,3), https://doi.org/10.1021/acsanm.2c05554
\bibitem{Mend}Mendonça, B., Moraes, E., Batista, R., Oliveira, A., Barbosa, M. \& Chacham, H. Water Diffusion in Carbon Nanotubes for Rigid and Flexible Models. {\em The Journal Of Physical Chemistry C}. \textbf{127}, 9769-9778 (2023), 
\bibitem{KOHLER201954}Köhler, M., Bordin, J., De Matos, C. \& Barbosa, M. Water in nanotubes: The surface effect. {\em Chemical Engineering Science}. \textbf{203} pp. 54-67 (2019), https://www.sciencedirect.com/science/article/pii/S0009250919303331
\bibitem{co2eGraf}Fuentes-Azcatl, R. \& Domínguez, H. Carbon Dioxide Confined between Two Charged Single Layers of Graphene: Molecular Dynamics Studies. {\em The Journal Of Physical Chemistry C}. \textbf{123}, 23705-23710 (2019), 
\bibitem{VALENCIAORTEGA2019243}Valencia-Ortega, M., Fuentes-Azcatl, R. \& Dominguez, H. Carbon dioxide adsorption on a modified zeolite with sodium dodecyl sulfate surfactants: A molecular dynamics study. {\em Journal Of Molecular Graphics And Modelling}. \textbf{92} pp. 243-248 (2019), https://www.sciencedirect.com/science/article/pii/S1093326319304632
\bibitem{tip4pe}Fuentes-Azcatl, R. \& Alejandre, J. Non-Polarizable Force Field of Water Based on the Dielectric Constant: TIP4P/$\epsilon$. {\em J. Phys. Chem. B}. \textbf{118}, 1263-1272 (2014,2)

\bibitem{spce}Fuentes-Azcatl, R., Mendoza, N. \& Alejandre, J. Improved SPC force field of water based on the dielectric constant: SPC/$\epsilon$. {\em Physica A: Statistical Mechanics And Its Applications}. \textbf{420} pp. 116-123 (2015), https://www.sciencedirect.com/science/article/pii/S0378437114009108
\bibitem{tip4pef}Fuentes-Azcatl, R. Flexible model of water based on the dielectric and electromagnetic spectrum properties: TIP4P/$\epsilon_{Flex}$. {\em Journal Of Molecular Liquids}. \textbf{338} pp. 116770 (2021), https://www.sciencedirect.com/science/article/pii/S016773222101494X
\bibitem{fbae}Fuentes-Azcatl, R. \& Barbosa, M. Flexible bond and angle, FBA/$\epsilon$ model of water. {\em Journal Of Molecular Liquids}. \textbf{303} pp. 112598 (2020), http://www.sciencedirect.com/science/article/pii/S0167732219331496
\bibitem{10.1063/1.468222}Chipot, C., Millot, C., Maigret, B. \& Kollman, P. Molecular dynamics free energy simulations: Influence of the truncation of long‐range nonbonded electrostatic interactions on free energy calculations of polar molecules. {\em The Journal Of Chemical Physics}. \textbf{101}, 7953-7962 (1994,11), https://doi.org/10.1063/1.468222
\bibitem{Van_Der_Spoel2005-oo}Van Der Spoel, D., Lindahl, E., Hess, B., Groenhof, G., Mark, A. \& Berendsen, H. GROMACS: fast, flexible, and free. {\em J Comput Chem}. \textbf{26}, 1701-1718 (2005,12)
\mciteBstWouldAddEndPuncttrue
\mciteSetBstMidEndSepPunct{\mcitedefaultmidpunct}
{\mcitedefaultendpunct}{\mcitedefaultseppunct}\relax
\EndOfBibitem
\bibitem{1995JChPh.103.8577E}Essmann, U., Perera, L., Berkowitz, M., Darden, T., Lee, H. \& Pedersen, L. A smooth particle mesh Ewald method. {\em jcp}. \textbf{103}, 8577-8593 (1995,11)
\mciteBstWouldAddEndPuncttrue
\mciteSetBstMidEndSepPunct{\mcitedefaultmidpunct}
{\mcitedefaultendpunct}{\mcitedefaultseppunct}\relax
\EndOfBibitem

\bibitem{10.1063/1.1378321}Tuckerman, M., Liu, Y., Ciccotti, G. \& Martyna, G. Non-Hamiltonian molecular dynamics: Generalizing Hamiltonian phase space principles to non-Hamiltonian systems. {\em The Journal Of Chemical Physics}. \textbf{115}, 1678-1702 (2001,7), https://doi.org/10.1063/1.1378321
\bibitem{je60064a005} Kell, G. Density, thermal expansivity, and compressibility of liquid water from 0.deg. to 150.deg.. Correlations and tables for atmospheric pressure and saturation reviewed and expressed on 1968 temperature scale. {\em Journal Of Chemical and Engineering Data}. \textbf{20}, 97-105 (1975)
\bibitem{10.1063/1.1461829}Wagner, W. \& Pruß, A. The IAPWS Formulation 1995 for the Thermodynamic Properties of Ordinary Water Substance for General and Scientific Use. {\em Journal Of Physical And Chemical Reference Data}. \textbf{31}, 387-535 (2002,6), https://doi.org/10.1063/1.1461829
\bibitem{2000CP}Soper, A. The radial distribution functions of water and ice from 220 to 673 K and at pressures up to 400 MPa. {\em Chemical Physics}. \textbf{258}, 121-137 (2000,8)
\bibitem{Neumann}Neumann, M. Dipole moment fluctuation formulas in computer simulations of polar systems. {\em Molecular Physics}. \textbf{50}, 841-858 (1983), 

\bibitem{PhysRevLett.98.247401}Sharma, M., Resta, R. \& Car, R. Dipolar Correlations and the Dielectric Permittivity of Water. {\em Phys. Rev. Lett.}. \textbf{98}, 247401 (2007,6), https://link.aps.org/doi/10.1103/PhysRevLett.98.247401

\end{mcitethebibliography}

\end{document}